\documentclass{PoS}
\usepackage{graphicx}
\usepackage{amssymb}
\usepackage{epstopdf}
\usepackage{subfigure}
\usepackage{graphicx}
\usepackage{soul}

\title{Chaotic Effects on Cosmic Ray Anisotropy in a Heliosphere-inspired Model} 

\ShortTitle{Chaotic Effects on Cosmic Ray Anisotropy in a Heliosphere-inspired Model}

\author{\speaker{Vanessa L\'opez-Barquero} \\
	Wisconsin IceCube Particle Astrophysics Center (WIPAC), University of Wisconsin, Madison, WI 53703, USA\\
	Department of Physics, University of Wisconsin, Madison, Wisconsin 53706, USA \\
	E-mail: \email{lopezbarquer@wisc.edu}
}

\author{Paolo Desiati\\
	      Wisconsin IceCube Particle Astrophysics Center (WIPAC), University of Wisconsin, Madison, WI 53703, USA\\
	       E-mail: \email{paolo.desiati@icecube.wisc.edu }}

\abstract{Cosmic rays propagate through the Galaxy and encounter systems that may trap them temporarily, as well as magnetic field structures that induce chaotic behavior on their trajectories. In particular, this is the case for particles that propagate in the local interstellar medium and interact with the heliospheric magnetic field before being detected on Earth. As a consequence, the observed cosmic-ray arrival direction distribution is affected by the heliosphere as long as their gyro-radius is smaller or comparable with the heliospheric size, i.e. in the TeV energy range. The chaotic nature of those cosmic-ray particle trajectories that are temporarily trapped inside the heliosphere can be characterized using the Finite-Time Lyapunov Exponents. Specifically, we will delve into the case of particles trapped in a heliospheric-inspired toy model of a static magnetic bottle configuration and the presence of temporal magnetic perturbations. In this work, we also suggest that a time-variability could prove to be important in the understanding of the TeV cosmic-ray anisotropy at Earth.
}

\FullConference{36th International Cosmic Ray Conference -ICRC2019-\\
		July 24th - August 1st, 2019\\
		Madison, WI, U.S.A.}

\begin{document}

\section{Introduction} 
\label{sec:intro}

Observations show that the cosmic ray flux on Earth is not perfectly isotropic but it is characterized by a small energy-dependent anisotropy in their arrival direction distribution (see~\cite{icecube, hawc, hawc-ic} and references therein). The origin of such an anisotropy is still to be understood, although a few models were proposed and studies carried out to address the observations in the TeV energy range (\cite{laza-desiati, desi-lazarian, schwadron, ming, lopez-barquero-helio}).

Cosmic ray anisotropy is most likely a convolution of different phenomena, such as the distribution of their sources in the Galaxy, the complex geometry and properties of the turbulent interstellar magnetic fields (see, e.g. \cite{giacinti, ahlers, LB16}), or definite coherent structures e.g. the heliosphere (see, e.g. \cite{desi-lazarian}). 

The interaction of cosmic rays with the heliosphere can have an enormous impact on the final arrival direction, especially on cosmic rays with rigidities between 1 and 10 TV. In this work, we will explore the possibility that particle trajectories may develop chaotic behavior while traversing and being temporarily trapped in the heliosphere. 

For this purpose, we have created a toy model that captures large-scale magnetic features of the heliosphere, such as the draping of the interstellar magnetic field and the solar cycles.

\section{Magnetic Field}
\label{sec:magf}

In the interest of assessing the effects that the heliosphere can have on cosmic rays, a toy model is constructed. Magnetic mirrors are used to simulate the influence on the arrival direction distribution produced by the draping of the interstellar medium magnetic field (see e.g. \cite{pogorelov}). Specifically, we studied the trapping properties of particles in this magnetic mirror. Such a toy magnetic field configuration is generated by employing two circular coils with electric currents running in the same direction, so to form an axially symmetric magnetic bottle. The  parameters used for its radius, distance, and current are chosen so that it will have characteristics similar to the ones present in the heliosphere. 
The radius and current produce a 3 $\mu$G field at the center of each coil and 1 $\mu$G between the two coils. The distance of the magnetic field line bending around the heliosphere is taken to be 2000 AU. 

In order to replicate the effects of solar cycles (see, e.g. \cite{pogorelov2}) in our toy model, we added a time-dependent perturbation propagating transversally through the magnetic bottle (along the x-axis). Periodic modulations along the y-axis and a gaussian dependency along the magnetic bottle axis are introduced, with the largest perturbation located at the center of the magnetic system. Such magnetic perturbation is represented by the function:

\begin{equation}
B_{y} = \frac{\Delta B}{B}\,\sin(k_p x-\omega_p t)\,e^{-\frac{1}{2}\left(\frac{z}{\sigma_p}\right)^2}
\label{eq:pert}
\end{equation}

where $k_p = \frac{2\pi}{L_p}$ and $\omega = \frac{2\pi v_p}{L_p}$ with $L_p$ = 200 AU the spatial scale of the magnetic polarity regions, $\sigma$ = 200 AU the width of the gaussian modulation of the perturbation. The relative amplitude $\frac{\Delta B}{B}$ and velocity $v_p$ depend on the strength and type of magnetic perturbation, as shown in table~\ref{tab:param}. An induced electric field, $\mathbf{E} = - \mathbf{v} \times \mathbf{B}$, has been added to one of the configurations. However, the magnitude of the electric force is approximately three orders of magnitude smaller, so no notable effects from it are expected.

While the magnetic bottle toy model is fundamentally different from the heliosphere, its purpose is to study the global properties of those particle trajectories where the adiabaticity limit is broken and that develop chaotic behavior. With this model, we developed and tested the diagnostics tools to investigate and characterize the chaotic properties of the cosmic ray trajectories that transverse the heliosphere. A follow up study will report the results obtained with a dynamical model of the heliosphere.

\begin{table}
\label{tab:b}
	\centering
	\caption{Parameters for the Time-Perturbation}
	\label{tab:param}
	\begin{tabular}{ | p{2.4cm} |  p{1.5cm} | p{1.5cm} |p{1.5cm} | }
		\hline \hline
		& Weak &Weak + E & Strong   \\ 
		\hline
		$\frac{\Delta B}{B}$ ($\mu$G)&  0.1 & 0.1 & 0.5\\ 
		\hline
		$v_p$ (AU/yr) & 2 & 2 & 20 \\
		\hline
	\end{tabular} 
\end{table}

\begin{figure}
	\centering
	\includegraphics[width=1\linewidth]{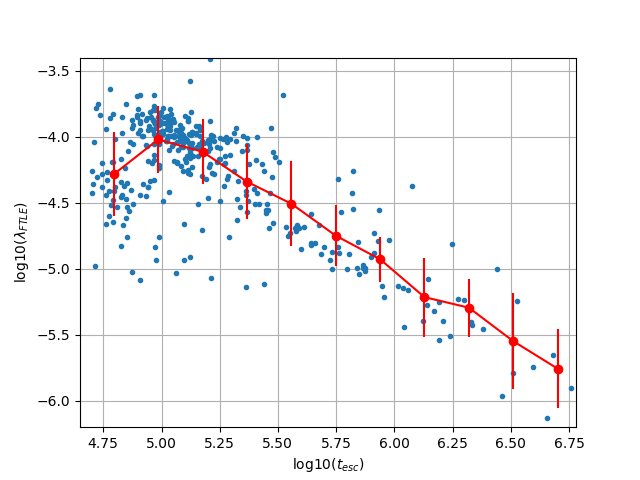}
	\caption[unperturbed]{ \textit{Unperturbed system}: The Finite-Time Lyapunov Exponent, $\lambda_{\tiny{FTLE}}$, vs. the escape time from the system, $t_{esc}$, for the unpertubed system. The blue points denote the specific values for each set of particles, which correspond to different initial conditions. The profile is denoted by the red points and red line connecting them. The vertical red error bars correspond to one standard deviation. Note that from $t_{esc}\sim10^5$ to the maximum escape time, the distribution follows a power law-like behavior. The fit for the power law of the profile is given by the Eq. \ref{eq:powerlaw}, with shows a power of -1.04.}
	\label{fig:finitelyapunovvsescapetimeunperturbedchi2500}
\end{figure}

\begin{figure*}
	\centering
	\includegraphics[width=1\linewidth]{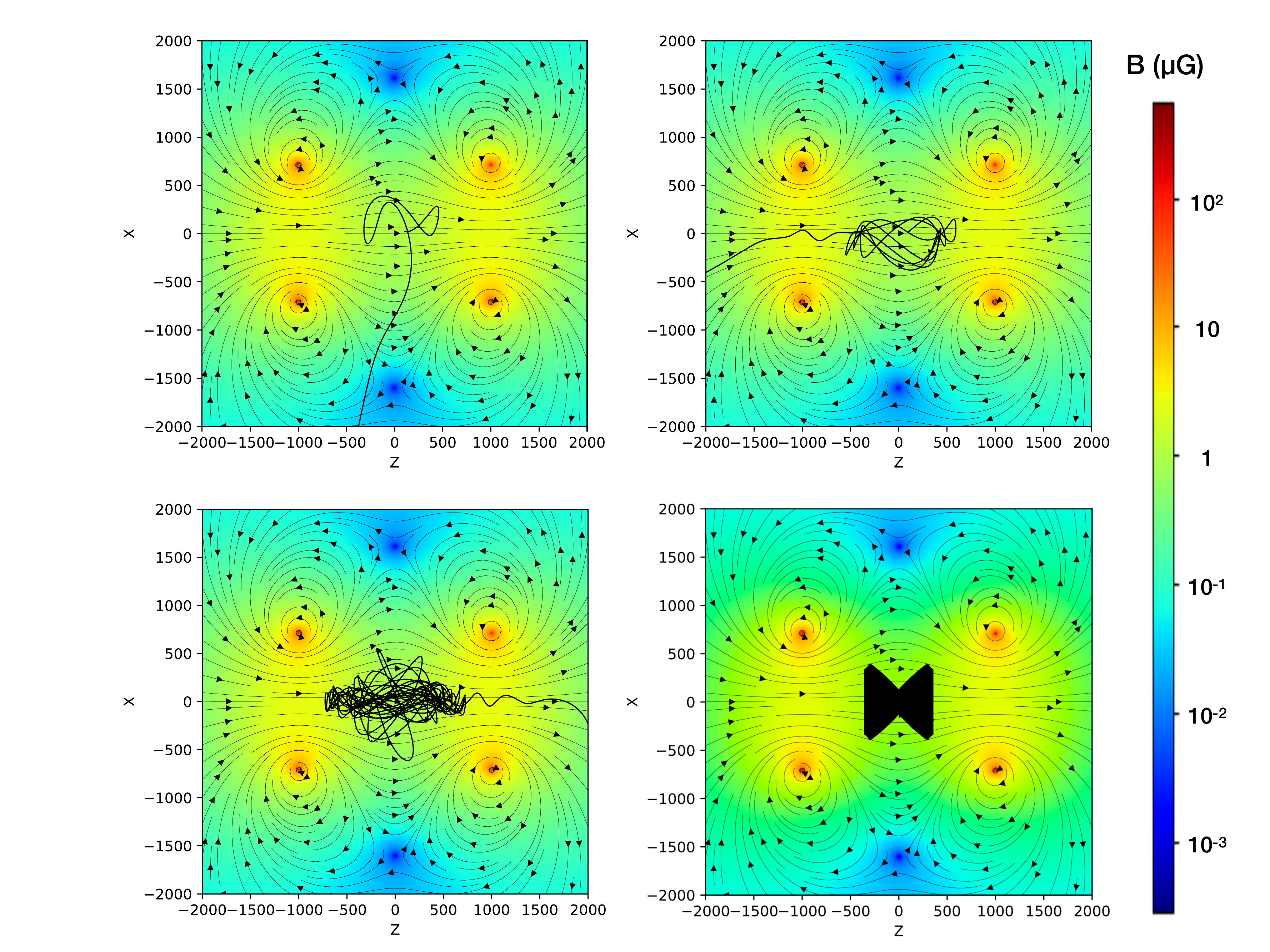}
	\caption[trajectories]{Trajectories in the unperturbed system. \textit{Top Left}: Transient particle with a escape time of 33000.  \textit{Top Right}: Intermediate particle with $t_{esc}=75402$  \textit{Bottom Left}: Particle in the power-law behavior section with a escape time of 295366. \textit{Bottom Right}: Trapped particle.}
	\label{fig:figure-trajectories}
\end{figure*}

\section{Cosmic Ray Trajectories}
\label{sec:crtraj}

Chaotic trajectories are very sensitive to the initial conditions. If these conditions change slightly, particles that originally started very close to each other can end up in completely different locations. One relevant feature is that these systems are still deterministic, however, there is an uncertainty on the initial conditions. This concept is also related to the observations since there is a limit on the resolution that can be achieved. Therefore, for a small solid angle in the sky, particles can be coming from vastly different original directions. In the systems described in this study, chaos comes from the geometry of the system and the intrinsic characteristics of the particles. Concretely, within one gyration and under certain conditions, particles may experience fast changes in the magnetic field.

One way to characterize chaotic trajectories is through the Lyapunov Exponents \cite{szezech, chirikov}. In this study, we utilize the Finite-Time Lyapunov Exponent (henceforth, FTLE) \cite{capsizing} since it provides more flexibility than the regular maximal Lyapunov Exponent. Specifically, we can deal with conditions in a bound system since a $t \rightarrow \infty $ condition is not necessary, and it is replaced by a finite time $\Delta t$.

The FTLE expression is given by: 
\begin{equation}\label{FTLE}
\lambda (t,\Delta t)=   \frac{1}{\Delta t}   \ln \left [ \frac{d(t+\Delta t)}{d(t) } \right ]
\label{eq:ftle}
\end{equation}
where the $\Delta t$ is the time interval for the calculation. The value for  $\Delta t$ is chosen depending on the intrinsic characteristics of the system and the particles traveling through it. 

This study is performed by numerically integrating the equation of motion for particle trajectories in the magnetic field described in Section \ref{sec:magf}. Four sets of trajectories were calculated, each with a different magnetic configuration. The first one, the unperturbed system, with just the static magnetic bottle configuration. Another set with the weak perturbation from eq.~\ref{eq:pert} (see table~\ref{tab:b}) added to the static magnetic bottle field. The third one with the weak perturbation and the induced electric field included (as described in sec.~\ref{sec:magf}). And a final one with the strong perturbation added (see table~\ref{tab:b}). For each set, a total of 768 anti-proton trajectories were integrated back in time, corresponding to each pixel in a HealPix grid \cite{healpix} with nside = 8. Then, for each of the 768 reference trajectories of the 4 sets, ten additional trajectories with the same initial momentum and with initial position randomly distributed around ($\hat{x}_0$, $\hat{y}_0$, $\hat{z}_0$) = (100, 100, 500) on a sphere of radius $\hat{r}_0$ = 0.01.

The method used for calculating the FTLE is based on the expression given by \ref{eq:ftle}. At each time step, the distance in phase space is calculated between each particle and its reference. With all the distances calculated, we proceed to the calculation of the Finite-Time Lyapunov Exponent. The election of the specific value for $\Delta t$ is done depending on the characteristics of the system, for example, in this case, the bouncing time between the magnetic mirror regions gives us a point of reference for the value of the $\Delta t$.  With these values of the FTLE obtained for each time step, a histogram is generated. We then proceed to fit two gaussians for the distribution. Since a positive value of the FTLE means divergence, we take the location of the gaussian with the highest positive expected value as our FTLE for each reference particle.

\section{Results}
\label{sec:res}

Based on our analysis for each reference particle and their correspondent set, we found that there is a correlation between the FTLE, i.e. their chaotic behavior, and the escape time from the system (Figure \ref{fig:finitelyapunovvsescapetimeunperturbedchi2500}) that follows a specific power-law.

\begin{figure}
	\centering
	\includegraphics[width=1\linewidth]{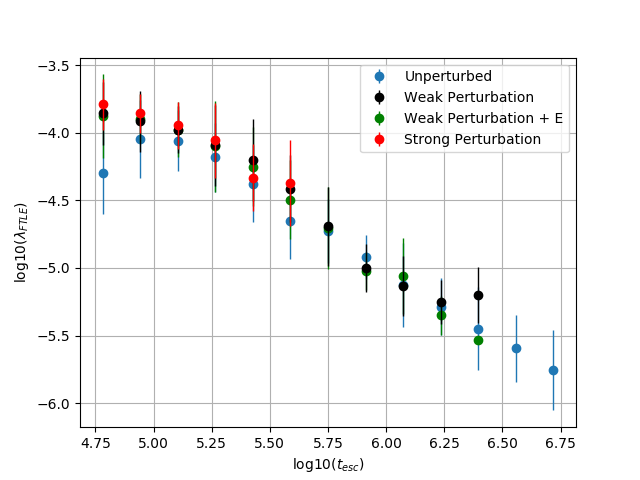}
	\caption[four cases]{  \textit{Comparison of perturbed systems.} The Finite-Time Lyapunov Exponent, $\lambda_{\tiny{FTLE}}$, vs. the escape time from the system, $t_{esc}$, for four different cases. The blue points represent the unperturbed system shown in figure \ref{fig:finitelyapunovvsescapetimeunperturbedchi2500}. The black points correspond to the profile of the weak-perturbation system, the green ones show the weak perturbation plus electric field, and the red ones the strong-perturbation system. Section \ref{sec:magf} shows the description for each magnectic field configuration. Note that once perturbations are introduced in the system, the overall distribution of particles in the different categories changes; nonetheless, the power law behavior and slope remains the same. }
	\label{fig:finitelyapunovvsescapetimefourcaseschi2500}
\end{figure}

Additionally, different regimes for the particles' behaviors are found to exist in these systems (see figure \ref{fig:figure-trajectories}). Particles with the shortest escape time, less than $t_{esc}=50000$ (in code units), leave the system right away and do not have time to develop any chaotic behavior. 
Intermediate particles, with $t_{esc}=50000$ to $10^5$, are so chaotic that the system can't hold their rapid divergence and they escape quickly from it. 
However, the majority of the particles exhibit a power-law behavior, see figures \ref{fig:finitelyapunovvsescapetimeunperturbedchi2500} and \ref{fig:finitelyapunovvsescapetimefourcaseschi2500}. 
The correlation between the FTLE and the escape time of the system follows a power law behavior described by: 
\begin{equation}
\lambda_{\tiny{FTLE}} = 10^{1.2} t_{esc}^{-1.04}
\label{eq:powerlaw}
\end{equation}

In the case of the unperturbed system, this is the only environment where there are particles that can be permanently trapped (given that the maximum integration time used in this work was set to $t=10^8$ in code units, which corresponds to about 330 years)

Another feature found in this study is the universality of the power-law behavior, shown in figure \ref{fig:finitelyapunovvsescapetimefourcaseschi2500}. So that the particles stay in the same power-law even when different perturbations act on it. For example, if particles in the unperturbed system are subjected to either the strong or weak perturbations, their FTLE will increase accordingly to the escape time given the behavior  in \ref{eq:powerlaw}.

\begin{figure*}
	\centering
	\includegraphics[width=1\linewidth]{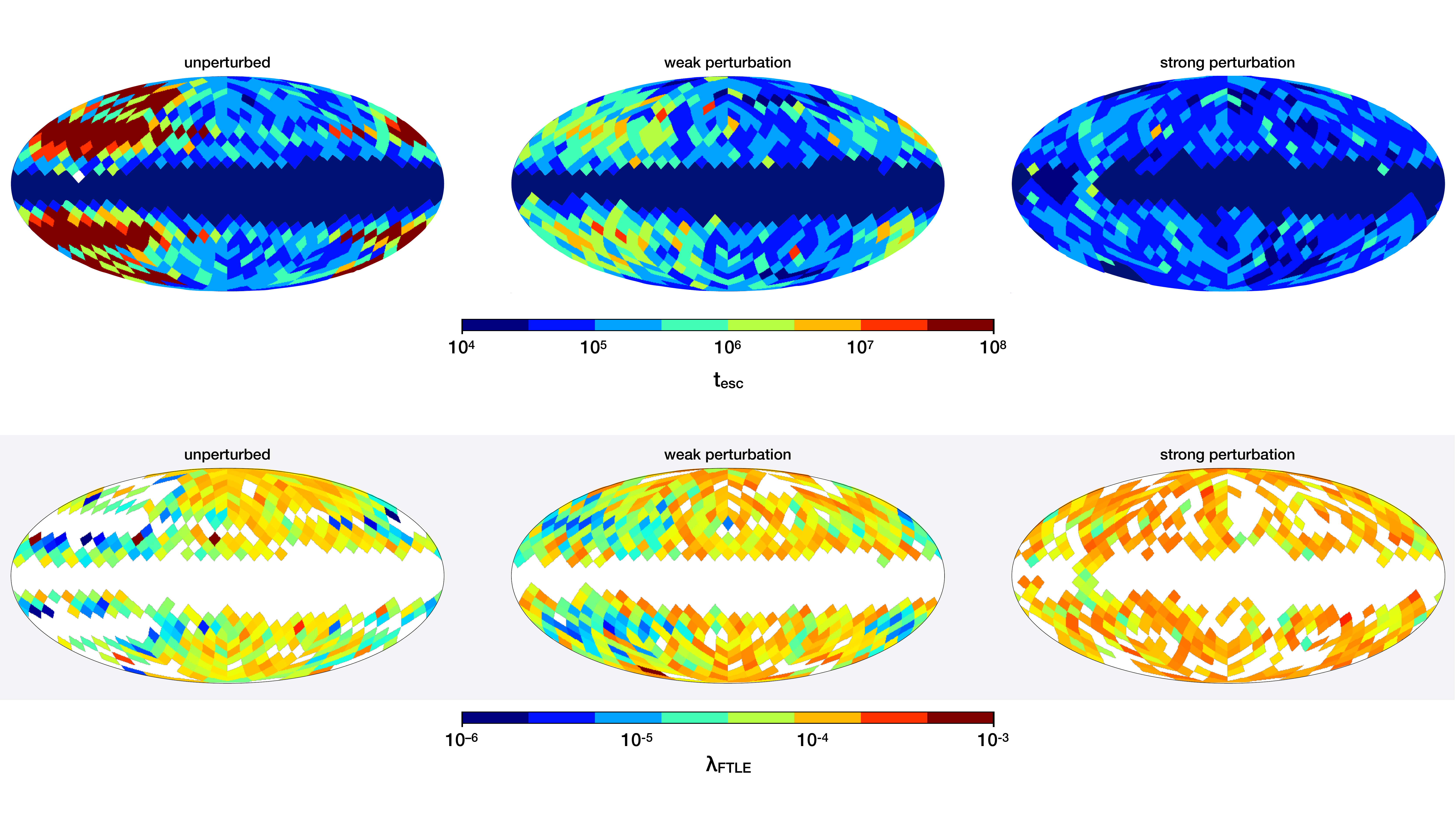}
	\caption[skymaps]{ \textit{Maps for arrival distribution.} The top panel corresponds to the escape times for the unperturbed, weakly-perturbed and strongly-perturbed system, respectively. The bottom panel corresponds to the Finite-Time Lyapunov Exponent, $\lambda_{\tiny{FTLE}}$ for those systems. }
	\label{fig:figure-skymaps}
\end{figure*}

Given the FTLEs and escape times calculated for the sets of trajectories, we plot them in an arrival distribution map. The location of each pixel corresponds to the arrival direction of a reference particle and the value for each pixel corresponds to either the FTLE or the escape time (see Figure \ref{fig:figure-skymaps}). In these maps we can identify that there are regions of stability where the particles are trapped, and originating from those regions, a gradient from longer to shorter escape times appear. Since $t_{esc}$ is related to the FTLE, the chaotic behavior of particles follows this trend as well.

\section{Discussion}
\label{sec:disc}

We have shown that particles in this heliospheric-inspired system display chaotic behavior. Such chaotic behavior can be characterized by the Finite-Time Lyapunov Exponent (FTLE), a quantity that can adapt to the specific bound conditions of our system. 
One important feature found in this study is that the FTLE is related to the escape time of the system. Its relation is a power-law given by the expression \ref{eq:powerlaw}, with the FTLE being approximately proportional to the inverse of the escape time. This power-law behavior is present even if perturbations act on the system, thus it points out to the possibility that this specific power can be used as a way to characterize the system. Additionally, if a system is found to have the same power, it is proposed that the consequences of the chaotic behavior will be similar to the ones shown in this heliospheric toy model. 

In a complementary manner, trajectories get more chaotic the stronger the perturbation that acts on the system, with the strongest perturbation being able to increase the FTLE by more than one order of magnitude, i.e. an increase in divergence of the trajectories. However, there is a maximum value of FTLE that particles can obtain in this system, $\lambda_{\tiny{FTLE}} \sim 10^{-4.0}$. This maximum FTLE is related to the size of the system, since the magnetic configuration is finite and can not contain a divergence larger than the one caused by this value.

The maps correspond to a visual representation for the distribution of chaotic behavior. In a scenario such as the heliosphere, trapped or particles with very long escape times most likely can not trace the distribution at a large distance, i.e. outside the heliosphere, since they have spent significant time inside and their memory of the initial distribution has been distorted. Nonetheless, transient trajectories, i.e. particles with the shortest escape times that do not have time to develop a chaotic behavior in the system, could provide a direct mapping between the initial and final configurations. 
Most of the particles in this system exhibit a chaotic behavior that can only provide an average mapping. Particles in this power-law regime are highly sensitive to the initial conditions, therefore, they are very likely to be affected by perturbations that could create a time-variability in the arrival distribution maps.

\section{Conclusions}
\label{sec:concl}

We have explored the possibility that chaotic behavior can be originated from the interaction between cosmic rays and a toy magnetic model inspired by the heliosphere, as well as the potential consequences that it can have on the cosmic ray arrival distribution. 

Our results show that the Finite-Time Lyapunov Exponent, a quantity that indicates the chaotic behavior of a trajectory, is related to the escape time of the system. This relation is given by a specific power-law that even persists if perturbations act on the system. The maps of arrival distribution display areas where the chaotic characteristics vary significantly, that could be used to predict time-variability in the observed cosmic ray anisotropy. 

Therefore, this study provides a basis for delving into these large-scale effects of the heliosphere on the TeV anisotropy, and it will open the possibility for a better understanding of the cosmic ray flux in the interstellar medium.

The study of the heliospheric effects on the distribution of the TeV cosmic rays, is important for the understanding of the global heliospheric properties (such as its dimension, the length of its tail, the scale of interstellar magnetic draping around the heliosphere, and so forth). But also for the unfolding of the pitch angle distribution of the cosmic rays as they approach the heliosphere, since it solely provides the information on the long time diffusion across the interstellar medium (see \cite{ming}). Chaotic trajectories have the loosest weight in the mapping between the observed directions and that in the interstellar medium, therefore it is important to make sure they are properly accounted for.

\end{document}